\documentclass[prd,aps,nofootinbib,preprint,tightenlines]{revtex4}

\newcommand{\checked}[1]{}

\usepackage{graphicx}

\newcommand{\beq}{\begin{equation}}
\newcommand{\eeq}{\end{equation}}
\newcommand{\bqa}{\begin{eqnarray}}
\newcommand{\eqa}{\end{eqnarray}}
\def\simge{\mathrel{
    \rlap{\raise 0.511ex \hbox{$>$}}{\lower 0.511ex \hbox{$\sim$}}}}
\def\simle{\mathrel{
    \rlap{\raise 0.511ex \hbox{$<$}}{\lower 0.511ex \hbox{$\sim$}}}}

\begin{document}

\title{The imaginary part of the static gluon propagator in an anisotropic
  (viscous) QCD plasma}
\preprint{BCCUNY-HEP/09-04}
\preprint{RBRC-783}

\author{Adrian Dumitru$^{a,b,c}$, Yun Guo$^{d,e}$, and
Michael Strickland$^{f}$}
\affiliation{$^a$Department of Natural Sciences, Baruch College, CUNY,
17~Lexington Ave, New York, NY 10010, USA\\
$^b$RIKEN-BNL Research Center, Brookhaven National Lab, Upton, NY
11973, USA\\
$^c$Graduate School and University Center, City University of New
York, 365 Fifth Avenue, New York, NY 10016, USA\\
$^d$Helmholtz Research School, Goethe Universit\"at,
Max-von-Laue-Str.\ 1, D-60438 Frankfurt am Main, Germany\\
$^e$Institute of Particle Physics, Huazhong Normal University,
Wuhan 430079, China\\
$^f$Department of Physics, Gettysburg College, Gettysburg, PA 17325, USA
\vspace*{2cm}
}

\begin{abstract}
  We determine viscosity corrections to the retarded, advanced and
  symmetric gluon self energies and to the static propagator in the
  weak-coupling ``hard loop'' approximation to high-temperature
  QCD. We apply these results to calculate the imaginary part of the
  heavy-quark potential which is found to be smaller (in magnitude)
  than at vanishing viscosity. This implies a smaller decay width
  of quarkonium bound states in an anisotropic plasma.
\end{abstract}
\maketitle
\newpage


{\em Introduction:} The retarded, advanced and symmetric gluon
self energy in an equilibrated, weakly-coupled high-temperature
plasma can be calculated within the ``hard thermal loop''
effective theory~\cite{{Braaten:1989mz},HTL}.  However, the
standard expressions receive corrections if the plasma undergoes
(anisotropic) expansion and if its shear viscosity is non-zero. In
this letter, we provide explicit expressions for the leading
viscous corrections to the retarded, advanced and symmetric
gauge-boson self energies and to the corresponding (static)
resummed propagators. As an application of particular interest, we
finally determine the corrections to the imaginary part of the
heavy-quark (static) potential in a QCD plasma. This imaginary
part provides a (contribution to the) width $\Gamma$ of quarkonium
bound states~\cite{Laine:2006ns,Laine:2007qy,Beraudo:2007} which
in turn determines their dissociation temperature: dissociation is
expected to occur when the binding energy decreases, with
increasing temperature, to $\sim\Gamma$~\cite{disso,disso2}. It is
therefore interesting to determine the effects of non-zero shear
viscosity on the imaginary part of the potential.

{\em Corrections to the thermal distributions in an anisotropic plasma:}
We consider a hot QCD plasma which, due to expansion and non-zero
viscosity, exhibits a local anisotropy in momentum space. The
phase-space distribution of particles is given by
\bqa
f({\bf p}) &=& f_{\rm iso}\left(\sqrt{{\bf p}^2+\xi({\bf p}\cdot{\bf
n})^2} \right) \\
&\approx& f_{\rm iso}(p) \left[ 1-\xi \frac{({\bf p}\cdot{\bf n})^2}
{2pT} \left( 1\pm f_{\rm iso}(p)\right)
\right]~.  \label{eq:f_aniso}
\eqa
Thus, $f({\bf p})$ is obtained from an isotropic distribution $f_{\rm
  iso}(|{\bf{p}}|)$ by removing particles with a large momentum
component along ${\bf{n}}$, the direction of
anisotropy~\cite{Romatschke:2003ms}. We shall restrict ourselves
here to a plasma close to equilibrium and so $f_{\rm iso}(p)$ is a
thermal ideal-gas distribution equal to either a Bose distribution
$n_B(p)$ or to a Fermi distribution $n_F(p)$, respectively.
Equation~(\ref{eq:f_aniso}) follows from an expansion in the
anisotropy parameter $\xi$. The correction $\delta f$ to the
equilibrium distribution exhibits precisely the structure expected
from viscous hydrodynamics for a fluid element expanding
one-dimensionally along the direction ${\bf n}$, provided we
identify~\cite{Asakawa:2006tc}
\beq \label{eq:xi_eta}
\xi = \frac{10}{T\tau} \frac{\eta}{s}~.
\eeq
Note that this expression only holds true in the Navier-Stokes
limit. In the general case, one can relate $\xi$ to the shear
tensor~\cite{Strickland:2009ce}. Here, $1/\tau$ denotes the
expansion rate of the fluid element and $\eta/s$ is the ratio of
shear viscosity to entropy density. As usual in viscous
hydrodynamics, the temperature $T$ as well as the entropy density
$s$ appearing in Eqs.~(\ref{eq:f_aniso}) and (\ref{eq:xi_eta}) are
defined at equilibrium; all viscous corrections are accounted for
explicitly by $\delta f$ in Eq.~(\ref{eq:f_aniso}).

{\em Propagators in the Keldysh real-time formalism:}
We shall calculate the finite-temperature self energies and
propagators using the well-known Keldysh real time
formalism~\cite{Carrington:1998jj}. The propagators are then $2 \times
2 $ matrices such as
\begin{eqnarray}
 \label{2a4}
  D(P)  =  \left (\begin{array}{cc} \frac{1}{P^2-m^2+i\epsilon} & 0\\
                             0 & \frac{-1}{P^2-m^2-i\epsilon}\\
            \end{array} \right ) -  2\pi i\, \delta (P^2-m^2)
 \left (\begin{array}{cc}
f_B & \Theta (-p_0)+f_B\\
\Theta (p_0)+f_B & f_B \\ \end{array} \right )
\end{eqnarray}
for a scalar field and
\begin{eqnarray}
S(P)=(\displaystyle{\not} P+m) && \left [ \left (\begin{array}{cc}
\frac{1}{P^2-m^2+i\epsilon} & 0\\
0 & \frac{-1}{P^2-m^2-i\epsilon}\\
                          \end{array} \right )\right .\nonumber \\
&& +2\pi i\, \delta (P^2-m^2)\> \left .\left (\begin{array}{cc}
f_F & -\Theta (-p_0)+f_F\\
-\Theta (p_0)+f_F & f_F \\ \end{array} \right )\right ],
\label{2a5}
\end{eqnarray}
for a Dirac field. We use the following notation: $P=(p_0,{\bf
p})$, $p=|{\bf p}|$. In equilibrium, the distribution functions
$f_B$ and $f_F$ correspond to Bose or Fermi distribution
functions, respectively. Away from equilibrium they need to be
replaced by the corresponding non-equilibrium distributions from
viscous hydrodynamics. It should be noted that Eqs.~(\ref{2a4})
and (\ref{2a5}) are ``bare'' propagators, the hard loop
resummation has yet to be performed.

The retarded, advanced and symmetric propagators can be obtained from
the Keldysh representation (which satisfies $D_{11}-D_{12}-D_{21}+D_{22}=0$) via
\begin{eqnarray}
\label{2a6}
   D_R = D_{11} - D_{12} ~,~ D_A = D_{11} - D_{21} ~,~
   D_F = D_{11} + D_{22}  ~.
\end{eqnarray}
In momentum space, the explicit expressions are
\begin{eqnarray}
D_R(P) & = & \frac{1}{P^2-m^2+i\, \mbox{sgn}(p_0) \epsilon},\nonumber \\
D_A(P) & = & \frac{1}{P^2-m^2-i\, \mbox{sgn}(p_0) \epsilon},\nonumber \\
D_F(P) & = & -2\pi i\, (1+2f_B)\, \delta (P^2-m^2) \label{2a10}
\end{eqnarray}
for scalar bosons and
\begin{eqnarray}
S_R(P) & = & \frac{\displaystyle{\not} P +m}{P^2-m^2+i\, \mbox{sgn}(p_0) \epsilon},\nonumber \\
S_A(P) & = & \frac{\displaystyle{\not} P +m}{P^2-m^2-i\, \mbox{sgn}(p_0) \epsilon},\nonumber \\
S_F(P) & = & -2\pi i\, (\displaystyle{\not} P +m)\, (1-2f_F)\, \delta (K^2-m^2)
\label{2a11}
\end{eqnarray}
for fermions, respectively.

In the real time formalism, similar relations hold for the self
energies:
\begin{eqnarray}
\Pi_{11}+\Pi_{12}+\Pi_{21}+\Pi_{22}=0 \label{2a12}
\end{eqnarray}
and
\begin{eqnarray}
\Pi_R  =  \Pi_{11}+\Pi_{12} ~,~
\Pi_A  =  \Pi_{11}+\Pi_{21} ~,~
\Pi_F  =  \Pi_{11}+\Pi_{22} ~. \label{2a13}
\end{eqnarray}

{\em Resummed photon (gluon) propagator:}
The resummed photon propagator can be determined from the
Dyson-Schwinger equation
 \begin{equation}
 \label{2b1}
   i{D^*} = i{D} + i{D} \, \bigl(-i\Pi\bigr)\, i{D^*} \, ,
 \end{equation}
where the propagators and self energy are $2\times 2$ matrices.
$D^*$ indicates a resummed propagator and $D$ is the bare
propagator.

Using the identities (\ref{2a6}) for the bare and resummed
propagators and (\ref{2a13}) for the self energies, it is easy to
show that
\begin{eqnarray}
 {D^*}_{R}=D_{R}+D_{R}\Pi_{R}{D^*}_{R}. \label{2b2}
\end{eqnarray}
A similar expression holds for the advanced propagator. The resummed
symmetric propagator satisfies
\begin{eqnarray}
 {D^*}_{F}=D_{F}+D_{R}\Pi _R{D^*}_{F}+D_F\Pi
_{A} {D^*}_{A}+D_{R}\Pi _{F}{D^*}_{A}~. \label{2b7}
\end{eqnarray}
Using $D_F(P) = (1+2f_B)\, \mbox{sgn}(p_0)\, [D_R(P)-D_A(P)]$
(this equation is true even in the
non-equilibrium case~\cite{Carrington:1998jj}), the solution for
${D^*}_{F}$ can be expressed in the form
\begin{eqnarray}
{D^*}_{F}(P)= && (1+2f_B)\, \mbox{sgn}(p_0)\,
[{D^*}_{R}(P)-{D^*}_{A}(P)]
\nonumber \\
&& +{D^*}_{R}(P)\,\{\Pi _F(P)-(1+2f_B)\, \mbox{sgn}(p_0)\, [\Pi
_R(P)-\Pi _A(P)]\} \,  {D^*}_A(P)~. \label{2b8}
\end{eqnarray}
In equilibrium we have $\Pi_F(P)=[1+2n_B(p_0)]\, \mbox{sgn}(p_0)\,
[\Pi _R(P)-\Pi _A(P)]$ and as a result the second term in
Eq.~(\ref{2b8}) vanishes. However, this is no longer true out of
equilibrium.

For the static potential we only require the temporal component of the
gluon propagator. At leading order in $\xi$ (or in the shear
viscosity) the calculation of ${D^*}^{00}$ is easiest in Coulomb
gauge. The temporal component decouples from the other components which
simplifies the calculation significantly. In this gauge (with the
gauge parameter $\eta=0$) the bare or resummed propagators satisfy
\begin{equation}
 K^i\cdot D^{0\,i}=0\,\,\, ,i=1,2,3
 \label{3b1}
\end{equation}
as a consequence of $\partial_{i} A^i=0$.
In the isotropic case this reduces to $D^{0\,i}=0$. We can then write
Eq.~(\ref{2b2}) as
\begin{eqnarray}
 {D^{*}}^ L_{R(0)}=D^L_{R}+D^L_{R}\Pi^L_{R(0)}{D^{*}}^
 L_{R(0)}~. \label{3b2}
\end{eqnarray}
Here, $L$ denotes the temporal component, $D^L\equiv D^{00}$.
A similar relation holds for the advanced propagator.
Eq.~(\ref{3b2}) no longer holds for the anisotropic system due
to breaking of isotropy in momentum space. However, for
small anisotropy, we can expand the resummed propagators and
self energies in $\xi$:
\begin{equation}
 D^*=D^{* }_{ (0)}+\xi D^{* }_{ (1)}+O(\xi^2) ,\,\,\,\,\,\,\,\Pi=\Pi_{(0)}+\xi
\Pi_{(1)}+O(\xi^2) ~.\label{3b4}
\end{equation}
The propagators to order $\xi^0$, $D^{* }_{ (0)}$ (either retarded,
advanced or symmetric) satisfy the equilibrium relations mentioned
above. For the linear term of order $\xi$,
\begin{equation}
 {D^* }_{R (1)}= D_R\, \Pi_{R (1)}{D^*}_{R (0)}+D_R\, \Pi_{R(0)}{D^*}_{R
 (1)}~.\label{3b51}
\end{equation}
For the temporal component, we have
\begin{equation}
 {D^* }^{L}_{R (1)}= (D_R)^{0\, \mu}\, ({\Pi_{R
     (1)}})_{\mu\,\nu}({D^*}_{R (0)})^{{\nu}\, 0}+(D_R)^{0\, \mu}\,
 ({\Pi_{R(0)}})
 _{\mu\,\nu}({D^*}_{R(1)})^{{\nu}\, 0}~.\label{3b52}
\end{equation}
Since $(D_R)^{0\,i}=0$, $({{D^*}_R}_{(0)})^{0\,i}=0$ and
$({\Pi_R}_{(0)})^{0\,i}\sim P^i$, according to Eq.~(\ref{3b1}), it
follows that\footnote{The temporal component of the retarded
propagator to
  order $\xi^2$, ${D^*}^ L_{R (2)}$, fails to satisfy such a
  relation. In fact, it includes a product of $(D_R)^{0\, \mu}\,
  (\Pi_{R (1)})_{\mu\,\nu}({D^*}_{R (1)})^{{\nu}\, 0}$, but
  $({\Pi_R}_{(1)})^{0\,i}$ is not proportional to $P^i$. As a result,
  $\sum\Pi_{R (1)}^{0\,i}{D^*}_{R (1)}^{i\,0}$ doesn't give zero
  automatically. This term will therefore depend on the spatial
  components of the self-energy and propagator which makes the
  calculation more complicated.}
\begin{equation}
 {D^*}^ L_{R (1)}= D^L_R\, \Pi_{R (1)}^L{D^*}^{ L}_{R (0)}+D_R^L\,
 \Pi_{R (0)}^L{D^*}^ L_{R (1)}~.\label{3b6}
\end{equation}
Again, a similar relation holds for the advanced propagator.
For the symmetric propagator, finally,
\begin{eqnarray}
{D^*}^L_{F(1)}(P)&= & (1+2f_{B(0)})\, \mbox{sgn}(p_0)\, [{D^*
}^L_{R(1)}(P)-{D^*}^L_{A(1)}(P)]\nonumber \\&+&2f_{B(1)}\,
\mbox{sgn}(p_0)\, [{D^* }^L_{R(0)}(P)-{D^*}^L_{A(0)}(P)]
\nonumber \\
&+&{D^*}^L_{R(0)}(P)\,\{\Pi _{F(1)}^L(P)-(1+2f_{B(0)})\,
\mbox{sgn}(p_0)\, [\Pi^L _{R(1)}(P)-\Pi^L _{A(1)}(P)]\ \nonumber
\\&-&2f_{B(1)}\,
\mbox{sgn}(p_0)\, [\Pi^L _{R(0)}(P)-\Pi^L _{A(0)}(P)]\} \,
{D^*}^L_{A(0)}(P)~.\nonumber\\ \label{3b7}
\end{eqnarray}

We now proceed to calculate explicitly the photon/gluon self energies
from which we obtain the propagators via the relations above. We
employ the diagrammatic Hard Loop approach but we have checked that
similar expression can be derived from Vlasov transport
theory~\cite{Mrowczynski:2000ed}. The contribution from the quark loop to the
gluon self energy is of the form
\begin{equation}
 \Pi^{\mu\nu}(P)=-\frac{i}{2}N_fg^2\int \frac
{d^4K}{(2\pi )^4} tr \left [\gamma ^\mu S(Q)\gamma^ \nu S(K)\right
], \label{3b8}
\end{equation}
where $S$ denotes the ``bare'' quark propagator and $Q=K-P$. Summing
the $11$ and $12$ components of the Keldysh representation leads to
\begin{eqnarray}
  \Pi
_R^L(P)=-i N_f g^2\int \frac{d^4K}{(2\pi )^4} (q_0k_0+{\bf q}\cdot
{\bf k}) && \biggl [\tilde \Delta _F(Q)\tilde \Delta _R(K)+\tilde
\Delta _A(Q)
\tilde \Delta _F(K)\nonumber \\
&& +\tilde \Delta _A(Q)\tilde \Delta _A(K)+\tilde \Delta
_R(Q)\tilde \Delta _R(K)\biggr ], \label{3b9}
\end{eqnarray}
where we neglect the fermion mass and write the
fermion propagator as $S_{R,A,F}(K)\equiv \displaystyle{\not} K \tilde \Delta
_{R,A,F}(K)$. The distribution function which appears in the
symmetric propagator has the form
\begin{equation}
f_F({\bf k}) = n_F(k)-\xi n_F^2(k) \frac{e^{k/T}}{2 k T} ({\bf
k}\cdot{\bf n})^2 + O(\xi^2)~,\label{3b10}
\end{equation}
with $n_F(p)$ a Fermi-Dirac function.
The last two terms of the integrand vanish after integration over
$k_0$. Temperature independent terms will be dropped in the
following. Shifting variables $K \to - K + P$ in the first term, we
found that the first two terms give the same contributions to the
final result. This is still true for a non-equilibrium distribution
which satisfies $f({\bf k})=f(-{\bf k})$. Then,
\begin{eqnarray}
  \Pi
_R^L(P)=4 \pi N_f g^2&&\int \frac{k d k d \Omega}{(2\pi )^4}
f_F({\bf k})
\biggl [ (2 k^2 - p_0 k-{\bf k}\cdot {\bf p})\frac{1}{P^2-2 k p_0+2{\bf k}\cdot {\bf p}-i\, \mbox{sgn}(k - p_0)\epsilon}\nonumber \\
&& +(2 k^2 + p_0 k-{\bf k}\cdot {\bf p})\frac{1}{P^2+2 k p_0+2{\bf
k}\cdot {\bf p}-i\, \mbox{sgn}(-k-p_0)\epsilon}\biggr ].
\label{3ba1}
\end{eqnarray}
Adopting the hard loop approximation, we assume that the internal
momenta are of order $T$ and therefore much larger than the
external momentum which is of order $gT$~\cite{Braaten:1989mz}.
The integrand in the square bracket can then be expanded in powers
of the coupling and the leading term is of the form
\begin{eqnarray}
\frac{2k^2}{-2kp_0+2{\bf k}\cdot {\bf p}-i\,
\epsilon}+\frac{2k^2}{2kp_0+2{\bf k}\cdot {\bf p}+i\, \epsilon}~.
\label{3ba2}
\end{eqnarray}
It can be easily shown that after integrating over $d\Omega$, this
contribution vanishes. The next to leading
term comes from the following 4 terms in the expansion of the
integrand in the square bracket
\begin{eqnarray}
\frac{-p_0k-{\bf k}\cdot {\bf p}}{-2kp_0+2{\bf k}\cdot {\bf p}-i\,
\epsilon}+\frac{p_0k-{\bf k}\cdot {\bf p}}{2kp_0+2{\bf k}\cdot
{\bf p}+i\, \epsilon}\nonumber\\-\frac{2k^2P^2}{(-2kp_0+2{\bf
k}\cdot {\bf p}-i\, \epsilon)^2}-\frac{2k^2P^2}{(2kp_0+2{\bf
k}\cdot {\bf p}+i\, \epsilon)^2}. \label{3ba3}
\end{eqnarray}
After integrating over $d\Omega$, we find that the first two terms
in Eq.~(\ref{3ba3}) give the same result; the last two terms also
contribute equally. As a result, the retarded self energy can be
expressed as
\begin{eqnarray}
  \Pi
_R^L(P)=\frac{4 \pi N_f g^2}{(2\pi)^4}\int k d k d \Omega f_F({\bf
k}) \frac{1-({\hat{\bf k}}\cdot {\hat{\bf p}})^2}{({\hat{\bf
k}}\cdot {\hat{\bf p}}+\frac{p_0+i\, \epsilon}{p})^2}.
\end{eqnarray}
Here, three-momenta with a hat denote unit vectors.

Using the distribution function~(\ref{3b10}) it is now
straightforward to calculate the retarded self energy:
\begin{eqnarray}
  \Pi
_R^L(P)=\frac{g^2}{2 \pi^2} N_f \sum\limits_{i=0,1} \int_{0}^{\infty} k\,
\Phi_{(i)}(k) d k \int_{-1}^{1} \Psi_{(i)}(s) d s ~, \label{3b11}
\end{eqnarray}
with
\begin{eqnarray}
\label{3b12}
\Phi_{(0)}(k)&=&n_F(k)\, ,\nonumber \\
\Phi_{(1)}(k)&=& - \xi n_F^2(k)\, \frac{e^{k/T} k}{2 T}\, ,\nonumber \\
\Psi_{(0)}(s)&=&\frac{1-s^2}{(s+\frac{p_0+i\epsilon}{p})^2}\, ,\nonumber \\
\Psi_{(1)}(s)&=& \cos^2\alpha\,
\frac{s^2(1-s^2)}{(s+\frac{p_0+i\epsilon}{p})^2}+\frac{\sin^2\alpha}{2}
\frac{(1-s^2)^2}{(s+\frac{p_0+i\epsilon}{p})^2}~.
\end{eqnarray}
Here, $\alpha$ is the angle between ${\bf{n}}$ and ${\bf{p}}$ and
$s\equiv {\hat{\bf k}}\cdot {\hat{\bf p}}$. This integral can be
performed analytically and the leading contribution for the isotropic
term is
\begin{equation}
 \Pi_{R(0)}^L(P)=\frac{g^2}{2 \pi^2} N_f \int_{0}^{\infty} k\,
\Phi_{(0)}(k) d k \int_{-1}^{1} \Psi_{(0)}(s) d s=N_f \frac{g^2
T^2}{6}\left (\frac{p_0}{2 p}\ln \frac{p_0+p+i\epsilon}
{p_0-p+i\epsilon} -1\right)~. \label{3b13}
\end{equation}
To order $\sim\xi$, the result is
\begin{eqnarray}
\Pi_{R(1)}^L(P)&=&\frac{g^2}{2 \pi^2} N_f \int_{0}^{\infty} k\,
\Phi_{(1)}(k) d k \int_{-1}^{1} \Psi_{(1)}(s) d s\nonumber \\
&=& N_f \frac{g^2 T^2}{6}\left(\frac{1}{6}+\frac{ \cos(2\alpha)}{2}\right)+
\Pi_{R(0)}^L(P)\left[\cos(2\alpha)-\frac{p_0^2}{2 p^2}(1+3
\cos(2\alpha))\right]~.\label{3b14}
\end{eqnarray}
In the Hard Loop limit the gluon-loop contributions to the gluon self
energy have the same structure as the one due to a quark loop.  (For
the same reason, the gluon self energy in the Hard Loop approximation
is gauge invariant.)  We can simply replace $N_f \frac{g^2 T^2}{6}$ by
$m_D^2$ to generalize the QED result to QCD. In the following, the
Debye mass $m_D^2$ is to be understood as
\begin{equation}
  m_D^2 = -{g^2\over 2\pi^2} \int_0^\infty d k \, k^2 \, {d
f_{\rm iso}(k) \over d k} ~. \label{3a13}
\end{equation}
This is independent of $\xi$ as all viscous corrections shall be
written explicitly. The isotropic distribution function
$f_{\rm iso}$ in our case is just a sum of the Fermi and Bose
distributions (with appropriate prefactors counting the number of
degrees of freedom~\cite{Romatschke:2003ms}). For $N_f$ massless quark
flavors and $N_c$ colors,
\begin{equation}
 m_D^2 = \frac{g^2 T^2}{6}(N_f+2 N_c)~. \label{3a14}
\end{equation}

Analogously, the advanced self energy is given by
\begin{equation}
\Pi_{A(0)}^L(P)=m_D^2\left (\frac{p_0}{2 p}\ln
\frac{p_0+p-i\epsilon} {p_0-p-i\epsilon} -1\right), \label{3b15}
\end{equation}
and
\begin{equation}
 \Pi_{A(1)}^L(P)= m_D^2\left(\frac{1}{6}+\frac{ \cos(2\alpha)}{2}\right)+
\Pi_{A(0)}^L(P)\left[\cos(2\alpha)-\frac{p_0^2}{2 p^2}\left(1+3
\cos(2\alpha)\right)\right]~. \label{3b16}
\end{equation}

Using Eqs.~(\ref{3b2}) and (\ref{3b6}), it is straightforward to
obtain the temporal component of the retarded propagator in
Coulomb gauge
\begin{equation}
{D^*}^L_{R(0)}=\left (p^2-m_D^2(\frac{p_0}{2 p}\ln
\frac{p_0+p+i\epsilon} {p_0-p+i\epsilon} -1)\right)^{-1}\,
,\label{3b17}
\end{equation}
\begin{equation}
{D^*}^L_{R(1)}=\frac{m_D^2(\frac{1}{6}+\frac{ \cos(2\alpha)}{2})+
\Pi_{R(0)}^L[\cos(2\alpha)-\frac{p_0^2}{2 p^2}(1+3
\cos(2\alpha))]}{(p^2-m_D^2(\frac{p_0}{2 p}\ln
\frac{p_0+p+i\epsilon} {p_0-p+i\epsilon} -1))^2}~.\label{3b18}
\end{equation}
Similar results can be obtained for the advanced propagator.
These results are identical to the
ones obtained within the transport theory approach.

Next we calculate $\Pi _F^L$ within the hard loop approximation.
Eq.~(\ref{2b8}) shows that this quantity is necessary to obtain
the resummed symmetric propagator out of equilibrium. Summing the
$11$ and $22$ components of the Keldysh representation, we obtain
\begin{eqnarray}
 \Pi _F^L(P)=-i N_f g^2\int
\frac{d^4K}{(2 \pi )^4}(q_0k_0+{\bf q}\cdot {\bf k})&&\biggl
[\tilde \Delta _F(Q)\tilde \Delta _F(K)-(\tilde \Delta _R(Q)
-\tilde \Delta _A(Q))\nonumber \\
&\times& (\tilde \Delta _R(K)-\tilde \Delta _A(K))\biggr ].
\label{3b19}
\end{eqnarray}
Using $\tilde \Delta _R(Q) -\tilde \Delta _A(Q)=-2\pi i\,
\mbox{sgn}(q_0)\delta(Q^2)$ and the hard loop approximation, we
find that the symmetric self energy can be expressed as
\begin{eqnarray}
 \Pi _F^L(P)=4 i N_f g^2 \pi^2\int
\frac{k^2dkd\Omega}{(2 \pi )^4}f_F({\bf k})(f_F({\bf
k})-1)\frac{2}{p}\biggl
[\delta(s+\frac{p_0}{p})+\delta(s-\frac{p_0}{p})\biggr ].
\end{eqnarray}
Again, we can expand the anisotropic distribution function to
order $\xi$ and finally arrive at
\begin{eqnarray}
 \Pi _{F (0)}^L(P)&=&-2 \pi i\,m_D^2 \frac{T}{p}\Theta(p^2-p_0^2)\, ,\nonumber \\
 \Pi _{F (1)}^L(P)&=& \frac{3}{2} \pi i\,m_D^2 \frac{T}{p}\left(\sin^2\alpha +
 (3 \cos^2\alpha -1 )\frac{p_0^2}{p^2}\right)\Theta(p^2-p_0^2)~. \label{3b20}
\end{eqnarray}

Next, we can calculate the symmetric propagator. We first consider the
isotropic case with $\xi=0$ and perform a Taylor expansion assuming
$p_0\to0$:
\begin{eqnarray}
{D^*}^L_{R(0)}(P)-{D^*}^L_{A(0)}(P)=\frac{m_D^2}{2p}\frac{-2\pi
i}{(p^2+m_D^2)^2}\, p_0~,
\end{eqnarray}
which follows from
\begin{eqnarray}
\lim_{p_0\to
0}(\ln\frac{p_0+p+i\,\epsilon}{p_0-p+i\,\epsilon}-
\ln\frac{p_0+p-i\,\epsilon} {p_0-p-i\,\epsilon})
=-2\pi i ~.
\end{eqnarray}
Similarly, when $p_0$ is small, the distribution function of on-shell
thermal gluons is
\begin{eqnarray}
(1+2n_B)\, \mbox{sgn}(p_0)= \frac{2 T}{p_0}~. \label{3ba4}
\end{eqnarray}
In the above equation, terms that do not contribute to the
symmetric propagator in the static limit have been neglected.
Finally, we can determine the temporal component of the symmetric
propagator explicitly in the isotropic limit:
\begin{equation}
 {D^*}^{L}_{F(0)}(p_0=0) = -\frac{2 \pi i T
m_D^2}{p\,(p^2+m_D^2)^2}~. \label{3b21}
\end{equation}
We now consider the contribution to order $\xi$. From
Eq.~(\ref{3b7}), the gluon distribution function can be expanded
as
\begin{eqnarray}
f_B= f_{B(0)}+\xi
f_{B(1)}=\frac{T}{|p_0|}-\frac{T\cos^2\alpha}{2|p_0|}\xi.
\label{3ba5}
\end{eqnarray}
There are 4 contributions at linear order of $\xi$ as shown in
Eq.~(\ref{3b7}). The calculation is similar to the isotropic case.
In addition, we need the following Taylor expansion to linear
order in $p_0$:
\begin{eqnarray}
{D^*}^L_{R(1)}(P)-{D^*}^L_{A(1)}(P)&=&-2\pi i
\left[\frac{m_D^4(1-3\cos(2\alpha))}{6
p(p^2+m_D^2)^3}+\frac{m_D^2\cos(2\alpha)}{2p(p^2+m_D^2)^2}\right]\,
p_0~,\nonumber \\
{\Pi^*}^L_{R(1)}(P)-{\Pi^*}^L_{A(1)}(P)&=&\frac{-\pi i
m_D^2\cos(2\alpha)}{p}\, p_0 ~.\label{3ba6}
\end{eqnarray}
Now the ${\cal O}(\xi)$ term of the symmetric propagator in the static limit
can be derived from the above equations
\begin{equation}
 {D^*}^{L}_{F(1)}(p_0=0) = \frac{3 \pi i T m_D^2}{2
p\,(p^2+m_D^2)^2}\sin^2\alpha -\frac{4 \pi i T m_D^4}{
p\,(p^2+m_D^2)^3}(\sin^2\alpha-\frac{1}{3})~. \label{3b22}
\end{equation}
%


{\em Heavy quark potential in an anisotropic plasma:} In the real
time formalism, the static heavy quark potential due to one gluon
exchange can be determined through the Fourier transform of the
physical ``11" component of the gluon propagator in the static
limit:
\begin{eqnarray}
V({\bf{r}},\xi) &=& -g^2 C_F\int \frac{d^3{\bf{p}}}{(2\pi)^3} \,
(e^{i{\bf{p \cdot r}}}-1)\left({D^*}^{L}(p_0=0,
  \bf{p},\xi)\right)_{11} \nonumber\\
&=& -g^2 C_F\int \frac{d^3{\bf{p}}}{(2\pi)^3} \, (e^{i{\bf{p \cdot
r}}}-1)\frac{1}{2}\left({D^*}^{L}_R+{D^*}^{L}_A+{D^*}^{L}_F\right)
\nonumber \\
&=& -g^2 C_F\int \frac{d^3{\bf{p}}}{(2\pi)^3} \, (e^{i{\bf{p \cdot
r}}}-1)\frac{1}{2}\left({D^*}^{L}_R+{D^*}^{L}_A\right)\nonumber\\
&-&g^2 C_F\int \frac{d^3{\bf{p}}}{(2\pi)^3} \, (e^{i{\bf{p \cdot
r}}}-1)\frac{1}{2}{D^*}^{L}_F ~. \label{41}
\end{eqnarray}
In the static limit, $\frac{1}{2}\left({D^*}^{L}_R +
{D^*}^{L}_A\right) = {D^*}^{L}_R = {D^*}^{L}_A$. The Fourier
transform of this quantity gives the real part of the potential
which has been discussed previously in
Refs.~\cite{Dumitru:2007hy,Dumitru:2009ni} and which determines
the quarkonium binding energies. Here, we instead consider the
imaginary part which comes from the Fourier transform of the
symmetric propagator. From Eqs.~(\ref{3b21}) and~(\ref{3b22}), the
isotropic contribution is given by
\begin{equation} {\bf{Im}}~V_{(0)}({{r}}) =-g^2 C_F\int
\frac{d^3{\bf{p}}}{(2 \pi)^3} (e^{i{\bf{p\cdot r}}}-1)\frac{- \pi
 T m_D^2}{p\,(p^2+m_D^2)^2} = - \frac{ g^2 C_F T}{4 \pi } \,
\phi(\hat{r})~, \label{46}
\end{equation}
with
\begin{equation}
 \phi(\hat{r})= 2\int_0^{\infty}dz \frac{
z}{(z^2+1)^2} \left[1-\frac{\sin(z\, \hat{r})}{z\, \hat{r}}\right]~,
\label{47}
\end{equation}
and $\hat{r}\equiv r \,m_D$. This result has been derived before
in Refs.~\cite{Laine:2006ns,Laine:2007qy}. The term of order $\xi$
can be expressed as
\begin{eqnarray}
 {\bf{Im} }~\xi V_{(1)}({\bf{r}}) &=& -g^2 C_F \, \xi \int
\frac{d^3{\bf{p}}}{(2 \pi)^3} \, (e^{i{\bf{p \cdot r}}}-1) \,\nonumber\\
&\times&\left[\frac{3 \pi  T m_D^2}{4
p\,(p^2+m_D^2)^2}\sin^2\alpha -\frac{2 \pi  T m_D^4}{
p\,(p^2+m_D^2)^3}(\sin^2\alpha-\frac{1}{3})\right]\,\nonumber\\
&=&\frac{ g^2 C_F \xi T}{4 \pi} \left[\psi_1(\hat{r},
\theta)+\psi_2(\hat{r}, \theta)\right]~,\label{48}
\end{eqnarray}
where $\theta$ is the angle between ${\bf{r}}$ and ${\bf{n}}$ and
\begin{eqnarray}
 \psi_1(\hat{r}, \theta) &=& \int_0^{\infty} dz
 \frac{z}{(z^2+1)^2}\left(1-\frac{3}{2}
 \left[\sin^2\theta\frac{\sin(z\, \hat{r})}{z\, \hat{r}}
 +(1-3\cos^2\theta)G(\hat{r}, z)\right]\right)\,~,\label{49}
\end{eqnarray}
\begin{eqnarray}
 \psi_2(\hat{r}, \theta) &=&- \int_0^{\infty} dz
\frac{\frac{4}{3}z}{(z^2+1)^3}\left(1-3 \left[
  \left(\frac{2}{3}-\cos^2\theta \right) \frac
 {\sin(z\, \hat{r})}{z\, \hat{r}}+(1-3\cos^2\theta)
 G(\hat{r},z)\right]\right)~,\label{410}
\end{eqnarray}
with
\begin{equation}
 G(\hat{r}, z)= \frac{\hat{r} z\cos(\hat{r} z)- \sin(\hat{r} z)
 }{(\hat{r} z)^3}~.
\label{411}
\end{equation}
\begin{figure}
\includegraphics[width=11.0cm]{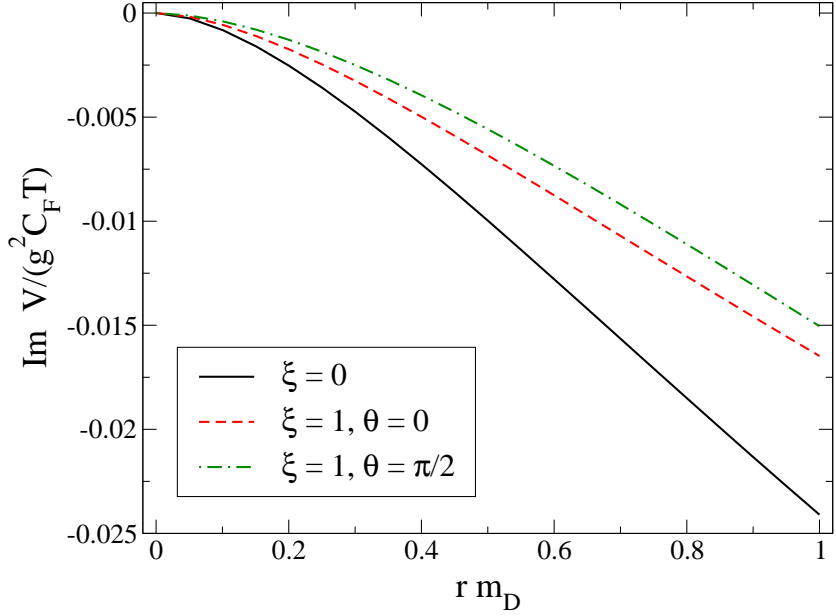}
\hspace{12mm} \caption[a]{Imaginary part of the static potential as a
  function of distance ($\hat{r}\equiv r\,m_D$). The vertical axis is
  ${\rm{Im}}~V/(g^2 C_F T)$. The curves, from bottom to top,
  correspond to an anisotropy of $\xi=0$ and $\xi=1$, $\theta=0$,
  $\theta={\pi}/{2}$.}
\label{fig:potential1}
\end{figure}
The result is shown in fig.~\ref{fig:potential1}. The imaginary part
decreases with $\xi$ (or with the viscosity, respectively).  When
$\hat{r}$ is small, we can expand the potential. This is relevant for
bound states of very heavy quarks whose Bohr radii $\sim1/(g^2 M_Q)$
are smaller than the Debye length $1/m_D$. For the imaginary part, at
leading order, the corresponding functions take the following forms:
\begin{eqnarray}
 \phi(\hat{r})&=& -\frac{1}{9}\, \hat{r}^2(-4+3\gamma_E+3\ln
 \hat{r})~,\nonumber \\
\psi_1(\hat{r}, \theta)&=&\frac{1}{600}\,
\hat{r}^2[123-90\gamma_E-90\ln
\hat{r}+\cos(2\theta)(-31+30\gamma_E+30\ln \hat{r})]~,\nonumber
\\\psi_2(\hat{r},
\theta)&=&\frac{1}{90}\,\hat{r}^2(-4+3\cos(2\theta))~,\label{51}
\end{eqnarray}
where $\gamma_E$ is the Euler-Gamma constant. At leading
logarithmic order then
\begin{eqnarray}
{\bf{Im}} ~V({\bf{r}},\xi) =-\frac{ g^2 C_F T}{4\pi}\,\hat{r}^2\,
\ln\frac{1}{ \hat{r}}
\left(\frac{1}{3}-\xi\frac{3-\cos(2\theta)}{20}\right)~.
\label{52}
\end{eqnarray}
Treating the imaginary part of the potential as a perturbation of the
vacuum Coulomb potential provides an estimate for the decay width,
\begin{eqnarray}
\Gamma &=&\frac{g^2 C_F T}{4\pi}\int d^3{\bf{r}}\,
\left|\Psi({{r}})\right|^2 \, \hat{r}^2\ln\frac{1}{
\hat{r}}\, \left(
\frac{1}{3}-\xi\frac{3-\cos(2\theta)}{20}\right)\nonumber\\
&=&\frac{16\pi T}{g^2
C_F}\frac{m_D^2}{M_Q^2}\left(1-\frac{\xi}{2}\right)\ln\frac{g^2C_FM_Q}{8\pi
m_D}~. \label{53}
\end{eqnarray}
Here, $M_Q$ is the quark mass and $\Psi({{r}})$ is the ground state
Coulomb wave function. Thus, at leading order in the deviation from
equilibrium (i.e., viscosity) the quarkonium decay width is smaller.
For a moderate anisotropy $\xi\simeq1$, $\Gamma$ decreases by about
50\% as compared to an ideal, fully equilibrated plasma.

{\em Acknowledgments:} We thank M.~Laine for useful comments on
the manuscript. Y.G.\ thanks the Helmholtz foundation and the Otto
Stern School at Frankfurt university for their support.

\vspace*{2cm} {\bf Note added:} While this paper was in the final
stages of preparation, another paper appeared where the imaginary
part of the heavy-quark potential at non-zero plasma anisotropy is
being considered~\cite{Burnier:2009yu}. The authors point out that
the ${\cal O}(\xi)$ correction contributes already at leading
non-trivial order to the quarkonium decay width, which agrees with
our finding. Their result differs from ours (numerically) but can
be reproduced if the second contribution on the r.h.s.\ of
Eq.~(\ref{2b8}) is omitted. From private communication with
M.~Laine, this appears to be due to a different setup of the
non-equilibrium system: Ref~\cite{Burnier:2009yu} considers a
situation where the soft gluons are in equilibrium at a
temperature $T$ while the hard gluons are out of equilibrium and
are characterized by a different hard scale $T'$. In contrast, we
assume that the deviation from equilibrium follows viscous
hydrodynamics, Eq.~(\ref{eq:f_aniso}).



\begin{thebibliography}{99}

\bibitem{Braaten:1989mz}
E.~Braaten and R.~D.~Pisarski,
Nucl.\ Phys.\  B {\bf 337}, 569 (1990).

\bibitem{HTL}
H.~A.~Weldon,
  Phys.\ Rev.\  D {\bf 26}, 1394 (1982);
J.~Frenkel and J.~C.~Taylor,
  Nucl.\ Phys.\  B {\bf 334}, 199 (1990);
M.~Le~Bellac, ``Thermal Field Theory'', Cambdridge University Press,
Cambridge, UK, 1996.

\bibitem{Laine:2006ns}
  M.~Laine, O.~Philipsen, P.~Romatschke and M.~Tassler,
  JHEP {\bf 0703}, 054 (2007)
  [arXiv:hep-ph/0611300].

\bibitem{Laine:2007qy}
  M.~Laine, O.~Philipsen and M.~Tassler,
  JHEP {\bf 0709}, 066 (2007)
  [arXiv:0707.2458 [hep-lat]].

\bibitem{Beraudo:2007}
 A.~Beraudo, J.~P.~Blaizot and C.~Ratti,
 Nucl.\ Phys.\  A {\bf 806}, 312 (2008) [arXiv:0712.4394
 [nucl-th]];
 N.~Brambilla, J.~Ghiglieri, A.~Vairo and P.~Petreczky,
 Phys.\ Rev.\  D {\bf 78}, 014017 (2008) [arXiv:0804.0993
 [hep-ph]].


\bibitem{disso}
A.~Mocsy and P.~Petreczky,
  Phys.\ Rev.\ Lett.\  {\bf 99}, 211602 (2007)
  [arXiv:0706.2183 [hep-ph]].

\bibitem{disso2}
Y.~Burnier, M.~Laine and M.~Vepsalainen,
  JHEP {\bf 0801}, 043 (2008)
  [arXiv:0711.1743 [hep-ph]];
M.~A.~Escobedo and J.~Soto,
  arXiv:0804.0691 [hep-ph];
M.~Laine,
  Nucl.\ Phys.\  A {\bf 820}, 25C (2009)
  [arXiv:0810.1112 [hep-ph]].

\bibitem{Romatschke:2003ms}
  P.~Romatschke and M.~Strickland,
  Phys.\ Rev.\  D {\bf 68}, 036004 (2003)
  [arXiv:hep-ph/0304092].

\bibitem{Asakawa:2006tc}
see, for example,
sections 5 and 6.6 in M.~Asakawa, S.~A.~Bass and B.~M\"uller,
Prog.\ Theor.\ Phys.\  {\bf 116}, 725 (2006)
[arXiv:hep-ph/0608270].

\bibitem{Strickland:2009ce}
 M.~Strickland and M. Martinez,
 arXiv:0902.3834 [hep-ph].

\bibitem{Carrington:1998jj}
M.~E.~Carrington, H.~Defu and M.~H.~Thoma,
Eur.\ Phys.\ J.\  C {\bf 7}, 347 (1999)
[arXiv:hep-ph/9708363];
Phys.\ Rev.\  D {\bf 58}, 085025 (1998)
[arXiv:hep-th/9801103].

\bibitem{Mrowczynski:2000ed}
  S.~Mrowczynski and M.~H.~Thoma,
  Phys.\ Rev.\  D {\bf 62}, 036011 (2000)
  [arXiv:hep-ph/0001164].

\bibitem{Dumitru:2007hy}
A.~Dumitru, Y.~Guo and M.~Strickland,
Phys.\ Lett.\  B {\bf 662}, 37 (2008)
[arXiv:0711.4722 [hep-ph]];
Y.~Guo,
Nucl.\ Phys.\  A {\bf 820}, 275C (2009)
[arXiv:0809.3873 [hep-ph]].

\bibitem{Dumitru:2009ni}
A.~Dumitru, Y.~Guo, A.~Mocsy and M.~Strickland,
arXiv:0901.1998 [hep-ph].

\bibitem{Burnier:2009yu}
Y.~Burnier, M.~Laine and M.~Vepsalainen,
arXiv:0903.3467 [hep-ph].

\end{thebibliography}
\end{document}